\documentclass[prb,twocolumn,superscriptaddress,preprintnumbers,amsmath,amssymb]{revtex4}
\usepackage{graphicx}
\usepackage{amsmath}
\usepackage{latexsym}
\usepackage{dcolumn}
\usepackage{bm}
\usepackage{float}
\usepackage{graphicx,here}

\emergencystretch=\maxdimen
\begin{document}
%\linenumbers
\title{Topological electronic structure in the antiferromagnet HoSbTe}

\author{Shaosheng Yue}
\affiliation{Institute of Physics, Chinese Academy of Sciences, Beijing 100190, China}
\affiliation{School of Physical Sciences, University of Chinese Academy of Sciences, Beijing 100049, China}
\author{Yuting Qian}
\affiliation{Institute of Physics, Chinese Academy of Sciences, Beijing 100190, China}
\affiliation{School of Physical Sciences, University of Chinese Academy of Sciences, Beijing 100049, China}
\author{Meng Yang}
\affiliation{Institute of Physics, Chinese Academy of Sciences, Beijing 100190, China}
\affiliation{School of Physical Sciences, University of Chinese Academy of Sciences, Beijing 100049, China}
\affiliation{Center of Materials Science and Optoelectronics Engineering, University of Chinese Academy of Sciences, Beijing 100049, China}
\author{Daiyu Geng}
\affiliation{Institute of Physics, Chinese Academy of Sciences, Beijing 100190, China}
\affiliation{School of Physical Sciences, University of Chinese Academy of Sciences, Beijing 100049, China}
\author{Changjiang Yi}
\affiliation{Institute of Physics, Chinese Academy of Sciences, Beijing 100190, China}
\affiliation{School of Physical Sciences, University of Chinese Academy of Sciences, Beijing 100049, China}
\affiliation{Center of Materials Science and Optoelectronics Engineering, University of Chinese Academy of Sciences, Beijing 100049, China}
\author{Shiv Kumar}
\affiliation{Hiroshima Synchrotron Radiation Center, Hiroshima University, 2-313 Kagamiyama, Higashi-Hiroshima 739-0046, Japan}
\author{Kenya Shimada}
\affiliation{Hiroshima Synchrotron Radiation Center, Hiroshima University, 2-313 Kagamiyama, Higashi-Hiroshima 739-0046, Japan}
\author{Peng Cheng}
\affiliation{Institute of Physics, Chinese Academy of Sciences, Beijing 100190, China}
\affiliation{School of Physical Sciences, University of Chinese Academy of Sciences, Beijing 100049, China}
\author{Lan Chen}
\affiliation{Institute of Physics, Chinese Academy of Sciences, Beijing 100190, China}
\affiliation{School of Physical Sciences, University of Chinese Academy of Sciences, Beijing 100049, China}
\affiliation{Songshan Lake Materials Laboratory, Dongguan, Guangdong 523808, China}
\author{Zhijun Wang}
\affiliation{Institute of Physics, Chinese Academy of Sciences, Beijing 100190, China}
\affiliation{School of Physical Sciences, University of Chinese Academy of Sciences, Beijing 100049, China}
\author{Hongming Weng}\thanks{hmweng@iphy.ac.cn}
\affiliation{Institute of Physics, Chinese Academy of Sciences, Beijing 100190, China}
\affiliation{School of Physical Sciences, University of Chinese Academy of Sciences, Beijing 100049, China}
\affiliation{Songshan Lake Materials Laboratory, Dongguan, Guangdong 523808, China}
\author{Youguo Shi}\thanks{ygshi@iphy.ac.cn}
\affiliation{Institute of Physics, Chinese Academy of Sciences, Beijing 100190, China}
\affiliation{School of Physical Sciences, University of Chinese Academy of Sciences, Beijing 100049, China}
\affiliation{Center of Materials Science and Optoelectronics Engineering, University of Chinese Academy of Sciences, Beijing 100049, China}
\affiliation{Songshan Lake Materials Laboratory, Dongguan, Guangdong 523808, China}
\author{Kehui Wu}
\affiliation{Institute of Physics, Chinese Academy of Sciences, Beijing 100190, China}
\affiliation{School of Physical Sciences, University of Chinese Academy of Sciences, Beijing 100049, China}
\affiliation{Songshan Lake Materials Laboratory, Dongguan, Guangdong 523808, China}
\author{Baojie Feng}\thanks{bjfeng@iphy.ac.cn}
\affiliation{Institute of Physics, Chinese Academy of Sciences, Beijing 100190, China}
\affiliation{School of Physical Sciences, University of Chinese Academy of Sciences, Beijing 100049, China}

\date{\today}

\begin{abstract}
{Magnetic topological materials, in which the time-reversal symmetry is broken, host various exotic quantum phenomena, including the quantum anomalous Hall effect, axion insulator states, and Majorana fermions. The study of magnetic topological materials is at the forefront of condensed matter physics. Recently, a variety of magnetic topological materials have been reported, such as Mn$_3$Sn, Co$_3$Sn$_2$S$_2$, Fe$_3$Sn$_2$, and MnBi$_2$Te$_4$. Here, we report the observation of a topological electronic structure in an antiferromagnet, HoSbTe, a member of the ZrSiS family of materials, by angle-resolved photoemission spectroscopy measurements and first-principles calculations. We demonstrate that HoSbTe is a Dirac nodal line semimetal when spin-orbit coupling (SOC) is neglected. However, our theoretical calculations show that the strong SOC in HoSbTe fully gaps out the nodal lines and drives the system to a weak topological insulator state, with each layer being a two-dimensional topological insulator. Because of the strong SOC in HoSbTe, the gap is as large as hundreds of meV along specific directions, which is directly observed by our ARPES measurements. The existence of magnetic order and topological properties in HoSbTe makes it a promising material for realization of exotic quantum devices.}
\end{abstract}

\maketitle

Topological quantum materials, such as topological insulators and semimetals, have attracted much attention in the last decade. These materials host topologically nontrivial electronic structures that can be characterized by symmetry-protected topological invariants \cite{HasanMZ2010,ArmitageNP2018,LvB2019}. Among the topological quantum materials, Dirac nodal line (DNL) semimetals \cite{FangC2016,YangSY2018,HuJ2019} are of particular interest because they are neighbors of various quantum states. In a DNL semimetal, the conduction and valence bands touch at extended lines in the three-dimensional Brillouin zone, and the degeneracy of the line nodes is protected by certain symmetries, such as mirror reflection, time-reversal, and spin-rotation symmetries. The breaking of these symmetries can fully or partially gap out the nodal lines, giving rise to rich topological quantum states, such as Dirac semimetals, Weyl semimetals, and topological insulators.

Prototypical DNL semimetals include the materials in the ZrSiS family \cite{SchoopLM2016,NeupaneM2016,ChenC2017,FuBB2019}. The existence of DNL fermions in these materials gives rise to various exotic properties, such as an extremely large and anisotropic magnetoresistance \cite{AliMN2016,LvYY2016,SinghaR2017}, a flat optical conductivity \cite{SchillingMB2017}, and an unconventional mass enhancement \cite{PezziniS2018}. Recently, LnSbTe (Ln=lanthanide) compounds \cite{XuQN2015,SinghaR2017prb,SchoopLM2018,HosenMM2018,LvB2019jpcm,WeilandA2019,PandeyK2020}, a special group in the ZrSiS family of materials, have received much attention because of the intrinsic magnetic order induced by certain lanthanide elements. The LnSbTe compounds have similar band structures to ZrSiS and host DNLs in proximity to the Fermi level. However, the DNLs of the ZrSiS family of materials only survive when the SU(2) spin-rotation symmetry is respected. When the spin-orbit coupling (SOC) effects are taken into account, the SU(2) spin-rotation symmetry is broken, which fully gaps out the DNLs. This drives the DNL semimetal states to weak topological insulator (WTI) states, with each layer being a two-dimensional (2D) TI \cite{QianYT2020}. In ZrSiS, the SOC effects are relatively weak, and the SOC gap is negligible. In the LnSbTe compounds, however, the strong SOC effects could result in a much larger gap, making realization of experimentally accessible WTI states possible. In addition, the existence of magnetic order in some LnSbTe compounds makes them an ideal platform to study the exotic properties of WTIs with broken time-reversal symmetry. However, direct experimental evidence of TI states in LnSbTe compounds has not yet been reported.

\begin{figure*}[htb]
\centering
\includegraphics[width=16 cm]{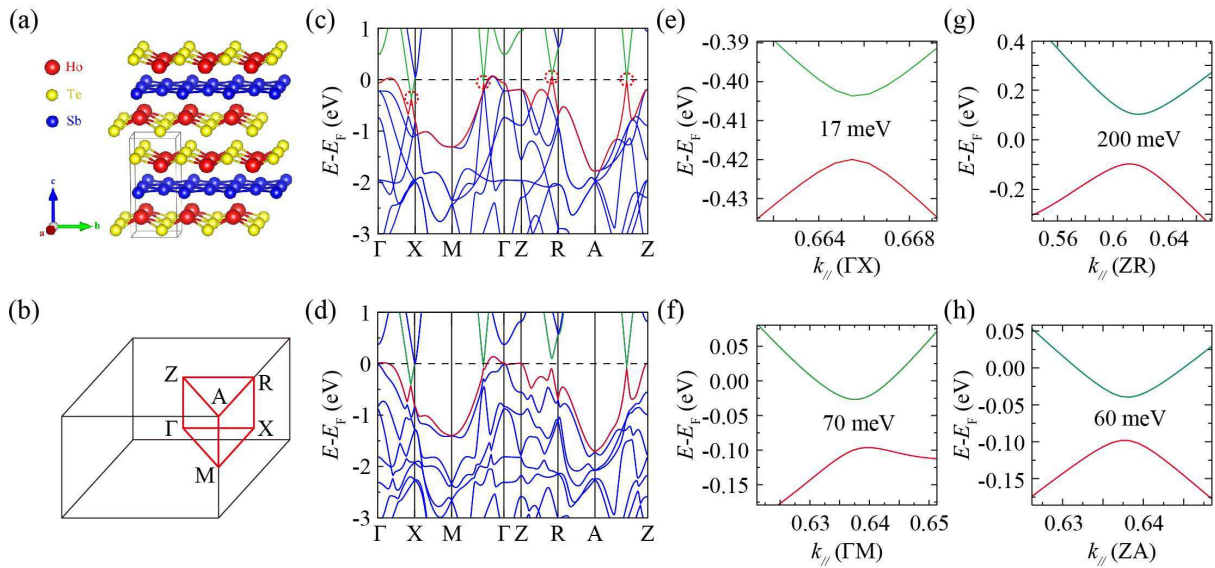}
\caption{(a) Crystal structure of HoSbTe. A unit cell is indicated by black lines. (b) Brillouin zone of HoSbTe. Red lines indicate high-symmetry directions. (c, d) Calculated bulk bands of HoSbTe along high-symmetry directions without and with SOC, respectively. Red and green lines in (c) highlight the highest valence and lowest conduction bands that form Dirac nodal lines. A schematic of all the Dirac nodal lines is shown in Fig. S1 of the Supplementary Materials. The red dotted circles in (c) indicate the degenerate points that form DNLs. (e-h) Calculated SOC-induced band gaps at the line nodes along high-symmetry directions: $\Gamma$--X, $\Gamma$--M, Z--R, and Z--A. The sizes of the gaps are indicated in the figure.}
\end{figure*}

Recently, single-crystal HoSbTe, a new member of the ZrSiS (or LnSbTe) family, was synthesized \cite{YangM2020}. Because of the unpaired 4f electrons of Ho atoms, HoSbTe is an antiferromagnetic (AFM) material with a $T_N$ of approximately 4 K. In this work, we performed angle-resolved photoemission spectroscopy (ARPES) and first-principles calculations to study the electronic structure and topological properties of HoSbTe. Our calculations show that HoSbTe is a DNL semimetal when SOC is neglected, analogous to other materials in the ZrSiS family \cite{SchoopLM2016,NeupaneM2016,ChenC2017,FuBB2019}. When SOC is taken into account, the otherwise gapless DNLs are fully gapped, giving rise to a paramagnetic WTI state. Our ARPES measurements show that along some directions, the gap can be as large as hundreds of meV, and the Fermi level is inside the gap.

High-quality single crystals of HoSbTe were grown by the Sb-flux method. High-purity Ho, Sb, and Te were mixed in an alumina crucible in a molar ratio of 1:20:1. The crucible was then placed in a quartz tube and sealed under a high vacuum atmosphere. The ampoule was heated to 1373 K over 15 h. After a dwell time of 10 h, the crucible was slowly cooled to 1073 K at a rate of 2 K/h, followed by centrifugation to separate the crystals from the excess Sb. ARPES measurements were performed at the BL-1 beamline of the Hiroshima Synchrotron Radiation Center \cite{IwsawaH2017}. The energy resolution of the ARPES measurements was approximately $\sim$15 meV. Fresh HoSbTe(001) surfaces for ARPES measurements were obtained by cleaving the crystals in ultrahigh vacuum while keeping the samples at low temperature. During the ARPES measurements, the temperature of the samples was kept at 30 K.

\begin{figure*}[htb]
\centering
\includegraphics[width=17cm]{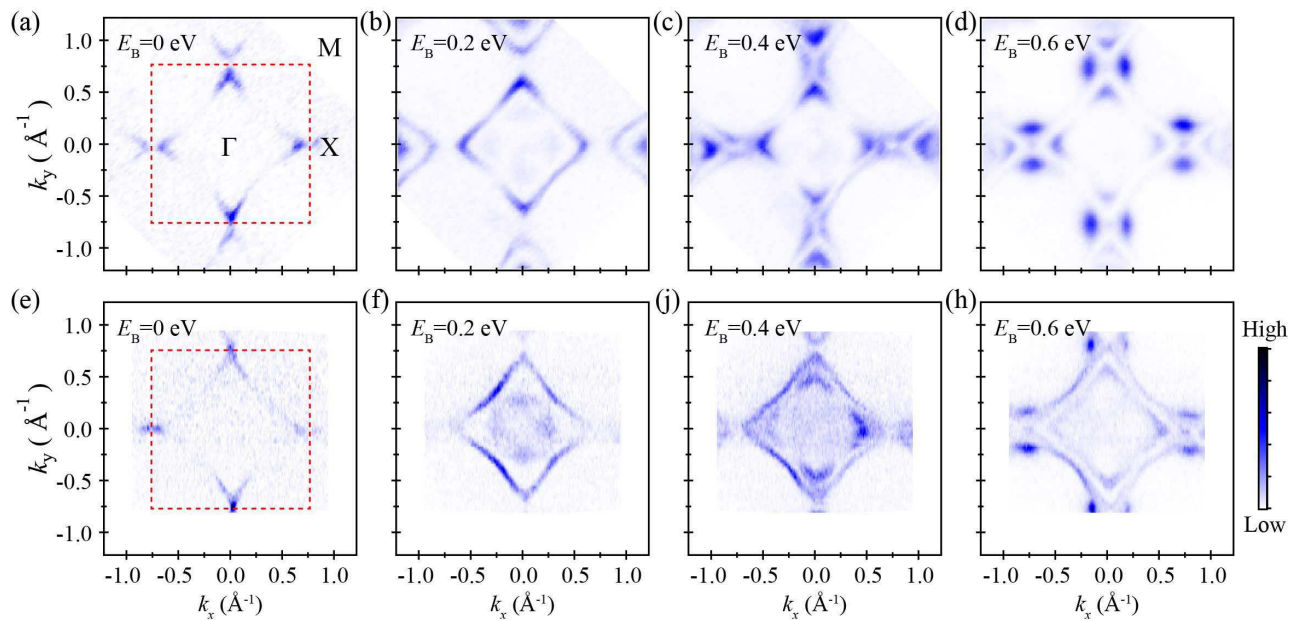}
\caption{ARPES intensity plots of the constant energy contours (CECs) of HoSbTe(001) at different binding energies: 0 eV, 0.2 eV, 0.4 eV, and 0.6 eV. (e)-(h) CECs measured with 120 eV photons. (a-d) CECs measured with 55 eV photons. Red dashed lines indicate the surface Brillouin zone of HoSbTe(001).}
\end{figure*}

First-principles calculations were performed using the Vienna {\it ab initio} simulation package (VASP) \cite{KresseG1996} based on density functional theory (DFT) on the basis of the projector augmented wave (PAW) \cite{BlochlPE1994,KresseG1999}. The generalized gradient approximation (GGA) of the Perdew-Burke-Ernzerhof (PBE) type \cite{PerdewJP1996} was adopted for the exchange-correlation potential between electrons. Because the experimental sample temperature ($\sim$ 30 K) was higher than the AFM transition temperature of HoSbTe ($T_N\sim$ 4 K), the magnetic order was not considered in our calculations. Since the magnetism was ignored and the 4f orbitals of Ho are far from the Fermi level ($\sim$ 6 eV), to simplify the calculation, the pseudopotential of Ho that does not contain 4f electrons was selected. A $k$-mesh sampling of 10$\times$10$\times$5 was adopted for self-consistent processes. The cutoff energy for the plane-wave basis was set to 300 eV. The band structures with and without SOC were obtained.

Single-crystal HoSbTe crystallizes in the nonsymmorphic space group P4/nmm (No. 129) with lattice constants $a$=$b$=4.23 \AA and $c$=9.15 \AA. The Ho-Te bilayers are sandwiched between square nets of Sb atoms, as shown in Fig. 1(a). The calculated bulk bands of HoSbTe without and with SOC are displayed in Figs. 1(c) and 1(d), analogous to other members of the ZrSiS family of materials \cite{XuQN2015}. When SOC is neglected, there are multipule DNLs in the 3D Brillouin zone of HoSbTe, as shown in Fig. S1 of the Supplementary Materials. Notably, a DNL exists at the $k_z$=0 and $\pi$ planes, and the degeneracy of the line nodes is protected by the glide mirror symmetry of the crystal. The degenerate points along high-symmetry directions are indicated by the red dotted circles in Fig. 1(c). When SOC is taken into account, the DNLs of HoSbTe are fully gapped, as shown in Fig. 1(d). To simplify the analysis of the topological properties of HoSbTe \cite{WatanabeH2018}, we neglected the magnetism arising from the 4f orbitals of Ho atoms. Our results show that HoSbTe becomes a WTI with symmetry-based indicator Z$_{2,2,2,4}$ =(0010), similar to the case of nonmagnetic LaSbTe \cite{QianYT2020}. The band topology of HoSbTe can also be understood by layer construction in real space \cite{SongZD2018}: each layer of HoSbTe is a 2D TI, and bulk HoSbTe can be viewed as a stacking of 2D TIs along the $c$ direction. Therefore, the symmetry-based indicator of HoSbTe can be fully obtained.

In ZrSiS, the SOC gap is negligible; therefore, ZrSiS is typically considered a DNL semimetal. In HoSbTe, however, the SOC is much stronger, which leads to a much larger gap than that in other members of the ZrSiS family. Figures 1(e)-1(h) show the calculated SOC gaps of HoSbTe. One can see that the gap reaches 200 meV along the Z--R direction, which is much larger than the thermal excitation energy. Such a large gap makes realization of the WTI states in HoSbTe possible in the experimentally accessible regime.

To confirm the intriguing electronic structure of HoSbTe, we performed high-resolution ARPES measurements on the (001) surface of HoSbTe. Because of the weak interlayer coupling, the ARPES spectra of HoSbTe show weak $k_z$ dispersion when the incident light is in the ultraviolet regime. This is analogous to other members of the ZrSiS family of materials \cite{SchoopLM2016,NeupaneM2016,ChenC2017,FuBB2019}. Figures 2(a)-2(d) and 2(e)-2(h) show the CECs of HoSbTe measured with 55 and 120 eV photons, which corresponds to $k_z$=12.3 and 17.2 $\pi$/$c$, respectively. Overall, the data at 55 eV and 120 eV are quite similar, despite the subtle differences such as the positions of the corners and intensities at the BZ boundary. One can identify a diamond-shaped Fermi surface (FS), with each apex pointing towards the $\bar{X}$ point. With increasing binding energy, the diamond-shaped FS splits into a pair: one shrinks towards the $\Gamma$ point, while the other one becomes larger.

The ARPES intensity of the band structure along the $\bar{\Gamma}$--$\bar{M}$ and $\bar{M}$--$\bar{X}$--$\bar{M}$ directions is displayed in Figs. 3(a)-3(d). For comparison, we present the slab calculation results along both directions in Figs. 3(e) and 3(f). The calculated band structures qualitatively agree with our experimental results. Along the $\bar{\Gamma}$--$\bar{M}$ direction, a parabolic band occurs that originates from the gapped DNL, as shown in Figs. 3(a)-3(c). The upper part of the gapped DNL is above the Fermi level and is thus inaccessible in our ARPES measurements. Notably, only one branch of the parabola along the $\bar{\Gamma}$--$\bar{M}$ direction can be observed with 34 and 55 eV photons, whereas both branches can be observed with 120 eV photons. The variation in the band intensity with the photon energy originates from the photoemission matrix element effects.

\begin{figure*}[t]
\centering
\includegraphics[width=16cm]{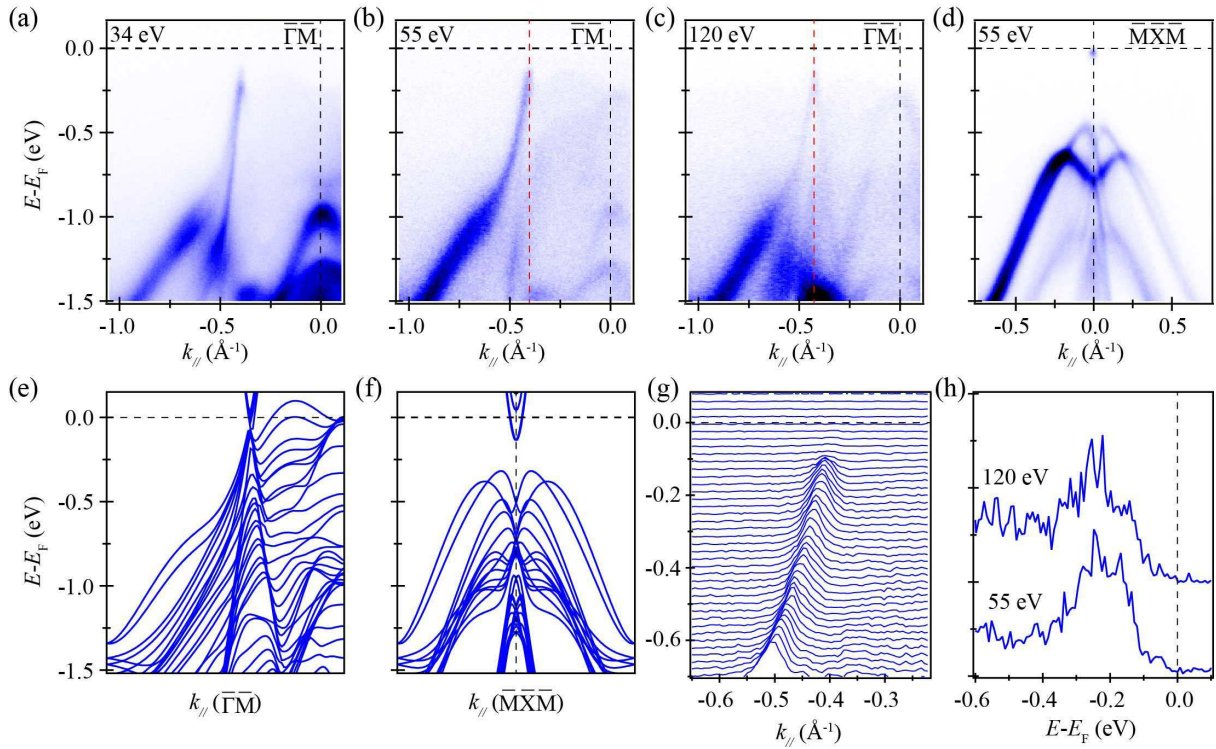}
\caption{(a)-(c) ARPES intensity plots of the band structure of HoSbTe along the $\bar{\Gamma}$--$\bar{M}$ direction measured with 34, 55, and 120 eV photons. Red dashed lines indicate the momentum position at which the energy distribution curves in (h) were taken. (d) ARPES intensity plots of the band structures along the $\bar{M}$--$\bar{X}$--$\bar{M}$ direction. (e, f) Slab calculation results of the band structure of HoSbTe along $\bar{\Gamma}$--$\bar{M}$ and $\bar{M}$--$\bar{X}$--$\bar{M}$, respectively. (g) Momentum distribution curve plots of (b). (h) Energy distribution curves along the red dashed lines in (b) and (c).}
\end{figure*}

Notably, the top of the parabolic band is well below the Fermi level, as shown in Figs. 3(a)-3(c). This feature is more evident in the momentum distribution curves in Fig. 3(g), which indicates the existence of an energy gap at the Fermi level. Figure 3(h) shows the energy distribution curve that passes through the top of the parabolic band, as indicated by the red dashed lines in Figs. 3(b) and 3(c). One can see that the band maximum along the $\Gamma$--M direction is located approximately 200 meV below the Fermi level. Since the upper part of the gapped nodal line is invisible in our ARPES measurements, the gap should be more than 200 meV. The size of the gap is much larger than the calculation results (Figs. 1(e)-1(h)), which might originate from the underestimation of band gaps in the DFT calculations. The opening of a large gap at the line nodes is also evident from the FS. As shown in Figs. 2(a) and 2(e), the spectral weight of the diamond-shaped FS is nonuniform: the spectral weight vanishes along the $\bar{\Gamma}$--$\bar{M}$ direction and recovers when approaching the $\bar{X}$ points, which indicates the existence of a band gap along the $\bar{\Gamma}$--$\bar{M}$ direction.

In summary, we report the observation of topological electronic structures in the antiferromagnetic material HoSbTe. When SOC is neglected, HoSbTe is a DNL semimetal, and its electronic structure is analogous to that of ZrSiS. However, the intrinsic strong SOC fully gaps out the nodal lines and drives the system to be a WTI. Each layer of HoSbTe is a 2D TI, and the bulk crystal can be viewed as a stacking of 2D TIs. The intriguing electronic structure of HoSbTe has been directly confirmed by ARPES measurements. Notably, the SOC gap of HoSbTe is as large as hundreds of meV, which is accessible for most experimental techniques. WTIs are quite rare, and only a few have been experimentally discovered, such as ZrTe$_5$ \cite{LiXB2016,WuR2016} and Bi$_4$I$_4$ \cite{NoguchiR2019}. The existence of magnetic order and WTI states makes HoSbTe a promising material for both fundamental research and device applications.

\begin{acknowledgments}
This work was supported by the Ministry of Science and Technology of China (Grant No. 2018YFE0202700), the National Natural Science Foundation of China (Grants No. 11974391, No. 11825405, and No. 1192780039), the Beijing Natural Science Foundation (Grant No. Z180007), and the Strategic Priority Research Program of the Chinese Academy of Sciences (Grants No. XDB33030100 and No. XDB30000000). The ARPES experiments were performed with the approval of the Proposal Assessing Committee of the Hiroshima Synchrotron Radiation Center (Proposal Numbers: 19AG006 and 19AG007).

%S.Y., Y.Q., M.Y., and D.G contributed equally to this work.
\end{acknowledgments}

\end{document}